\documentstyle[aaspp4]{article}
\def\plotone#1{\centering \leavevmode
\epsfxsize=\textwidth \epsfbox{#1}}

\begin{document}

\def\kms{km~s$^{-1}$}
\def\msun{$M_{\odot}$}
\def\rsun{$R_{\odot}$}
\def\lsun{$L_{\odot}$}
\def\halpha{H$\alpha$}
\def\hbeta{H$\beta$}
\def\hgama{H$\gamma$}
\def\hdelta{H$\delta$}
\def\Teff{T$_{eff}$}
\def\logg{$log_g$} 

\tighten

\received{25 April 2000}
\accepted{25 May 2000}
\journalid{AJ}{}
\slugcomment {Astronomical Journal, in press } 

\title{A Survey for Faint Stars of Large Proper Motion Using 
Extra POSS~II Plates\footnotemark[1]}

\author{David G. Monet}
\affil{U. S. Naval Observatory}  
\affil{Flagstaff, Arizona 86002}
\affil{dgm@nofs.navy.mil} 

\author{Matthew D. Fisher and James Liebert}
\affil{Steward Observatory}  
\affil{University of Arizona, Tucson AZ 85721}
\affil{matt19@worldnet.att.net,liebert@as.arizona.edu}

\author{Blaise Canzian and Hugh C. Harris}
\affil{U. S. Naval Observatory}  
\affil{Flagstaff, Arizona 86002}
\affil{blaise@nofs.navy.mil,hch@nofs.navy.mil} 

\author{and}

\author{I. Neill Reid}
\affil{Dept. of Physics \& Astronomy}
\affil{University of Pennsylvania} 
\affil{209 S. 33rd St.}
\affil{Philadelphia, PA 19104-6396}
\affil{inr@herschel.physics.upenn.edu} 

\footnotetext[1]{Portions of the data presented here were
obtained at the W.M. Keck Observatory, which is operated as a scientific
partnership among the California Institute of Technology, the University
of California, and the National Aeronautics and Space Administration.
The Observatory was made possible by the generous financial support of
the W.M. Keck Foundation.}


\begin{abstract}

We have conducted a search for new stars of high proper motion
($\geq$0.4 arcsec~yr$^{-1}$) using POSS~II fields for which an extra
IIIaF (red) plate of good quality exists, along with useable IIIaJ
(blue) and IV-N (infrared) plates taken at epochs differing by a minimum
of 1.5 years.  Thirty-five fields at Galactic latitudes $|b| \geq$
20$^o$ were measured, covering some 1378 deg$^2$, or 3.3\% of the sky.
Searches with three plate combinations as well as all four plates were
also made.  Seven new stars were found with $\mu\ \geq$0.5
arcsec~yr$^{-1}$, which were therefore missed in the Luyten Half Second
(LHS) Catalog.  One of these is a common proper motion binary consisting
of two subdwarf M stars; another is a cool white dwarf with probable
halo kinematics.  As a test of our completeness -- and of our ability to
test that of Luyten -- 216 of 230 catalogued high proper motion stars were
recovered by the software, or 94\%.  Reasons for incompleteness of the
LHS are discussed, such as the simple fact that POSS~II plates have
deeper limiting magnitudes and greater overlap than did POSS~I.
Nonetheless, our results suggest that the LHS is closer to 90\% complete
than recent estimates in the literature (e.g 60\%), and we propose a
reason to account for one such lower estimate.  The conclusion that the
LHS Catalog is more complete has implications for the nature of the halo
dark matter.  In particular it strengthens the constraint on the local
density of halo stars, especially white dwarfs at M$_V\sim$17-18. 

\end{abstract}


\section{INTRODUCTION}

Proper motion catalogs that include faint stars have generally been used
to obtain samples of low mass main sequence stars and white dwarfs of
both disk and halo populations in the solar neighborhood.  That this
kind of selection introduces both kinematic and population biases
(usually in favor of high velocity stars) has long been recognized
(cf. Hanson 1983).  The cumulative distribution function N($\mu$) of
a proper motion catalog should be proportional to $\mu^{-3}$, if the
catalog is complete to proper motion and magnitude limits (Luyten 1963,
Hanson 1979).  If, in addition, the kinematical properties of the
sample of stars being selected may be modelled accurately enough, the
bias can be removed in principle, and a luminosity function (LF) and
space density can be determined for the sample.

Schmidt (1975) first applied his 1/V$_{max}$ method to determine the LF
of local halo stars, using primarily stars from the proper motion
catalog of the Lowell Observatory (Giclas, Burnham and Thomas
1968). Bahcall \& Casertano (1986) applied a more comprehensive
kinematical model for halo stars to a similar sample drawn mainly from
Lowell proper motion measurements.  However, the Luyten Palomar survey
utilizing the Palomar Observatory Sky Survey (POSS~I) with the 48-inch
Schmidt Telescope in the 1950s and a second epoch series of red-only
(103aE) plates obtained by Luyten a decade later extended several
magnitudes fainter than the Lowell catalogs.  Using a sample of low mass
halo main sequence stars drawn largely from the stars of largest motion
in the Luyten Half Second (LHS) Survey (Luyten 1979), Dahn et al. (1995)
applied Schmidt's and Bahcall \& Casertano's methods to determine a
local halo LF extending to the stellar mass terminus.  Liebert, Dahn \&
Monet (1988, 1989) used a similar sample to estimate an LF for white
dwarf stars of the disk and a sampling of the halo.  Finally, there have
been several estimates of the disk low mass star LFs, and controversy
has arisen as to why estimates based on nearby star or proper motion
samples differ from those based on samples selected by red colors (Reid
and Gizis 1997).

The completeness of proper motion catalogs, especially for faint Luyten
stars of large motion, has also been a topic of some controversy.
Hanson's (1979) analysis suggested that the LHS Catalog may be 20\%
incomplete for proper motions $\mu<$1 arcsec~yr$^{-1}$. The analysis of
Dawson (1986) suggested that the LHS Catalog is 90\% complete to B= 21
over the region of sky north of $\delta$ = -30$^o$, above Galactic
latitude $|b|$ = 10$^o$.  More negative assessments of the Luyten
catalogs include those of Oswalt et al. (1996) and Wood \& Oswalt
(1998).  These authors applied a very large correction factor for
incompleteness of the Luyten catalogs, resulting in a higher space
density or LF for white dwarfs than was obtained by Liebert et
al. (1988, 1989) who assumed no incompleteness.  However, Wood and
Oswalt's (1998) simulations actually indicated that serious
magnitude-limited incompleteness sets in at $\mu_{lim}$ = 0.15-0.20
arcsec~yr$^{-1}$, and do not obviously apply to the motions of the
LHS. Although their conclusion was based on the discovery of a single,
cool white dwarf star (WD0346+246), Hambly, Smartt, \& Hodgkin (1997)
also inferred that the Luyten samples are incomplete.  Flynn et
al. (2000) concluded that the LHS is only 60\% complete to a red
magnitude (R$_L$) = 18.5, based on a statistical analysis of the New
Luyten Two Tenths (NLTT) Catalog.  It is worth pointing out again that
this was not a direct analysis of larger LHS motion stars. 

Of particular interest recently is the search for halo white dwarfs in
the solar neighborhood where they might be bright enough to be
detectable.  This interest has intensified due to the possible discovery
of faint stellar objects with proper motions in the Hubble Deep Field
(Ibata et al. 2000), and the possibility that MACHO collaboration
detections could be halo objects of around half a solar mass (Alcock et
al. 1997, 2000).  Some researchers have suggested that cool white dwarfs
could thus account for a substantial fraction of the dark halo or
corona.  The search for dim halo stars underscores another potential
shortcoming of the Luyten surveys, namely that the search radius of
Luyten's measuring machine limited the maximum detectable proper motion
to about 2.5$^{''}$~yr$^{-1}$.  As pointed out by Graff, Laughlin and
Freese (1998), this limited any search for nearby halo white dwarfs to
those with v$_{tan} \leq$121 km~s$^{-1}$ at 10pc distance.  Yet a
majority of halo white dwarfs nearby enough to the Sun to be found on
Schmidt plates may have higher tangential velocities than this. It
should nonetheless also be noted that Luyten personally blinked over
15\% of the POSS fields and found no stars with m$_R >$ 15 and $\mu >$
2.5$^{''}$~yr$^{-1}$. 

The Second Palomar Sky Survey (POSS~II) with the same telescope, now
called the Oschin Schmidt (Reid et al. 1991), introduces a valuable new
dataset to test the completeness of the earlier survey, and indeed to
find new proper motion stars to fainter limiting magnitudes (B$_J$ =
22.5 mag, R$_c$ = 20.8 I$_c$ = 19.5) due to the use of a new generation
of Kodak photographic emulsions and other improvements.  The
near-infrared bandpass had no counterpart in POSS~I.  In contrast
Luyten's ``second epoch'' for Palomar catalogs was obtained with the
same red Kodak emulsion type as POSS~I, and the exposure times were
shorter than for POSS~I.  In addition, the telescope was upgraded
considerably for the new survey, notably by replacing the corrector to
improve the images at near-infrared wavelengths, refining the mechanical
accuracy of the plate-holders and automating guiding.  Finally, for the
new survey a $5^{\circ}$ spacing between fields was adopted, instead of
the 6$^{\circ}$ spacing of POSS~I, the latter having barely allowed
overlap between adjacent fields.

The U.S. Naval Observatory, Flagstaff Station, has obtained a new
Precision Measuring Machine (PMM) for the purpose of digitizing all
POSS~I and POSS~II plates, as well as U.K. Science Research Council
SRC-J survey plates, and the European Southern Observatory ESO-R
survey plates, of generally similar quality as POSS for the Southern
Hemisphere.  The PMM design, operation, and data products are described
in documentary files on the USNO-A CD-ROM set and on the web site
http://www.nofs.navy.mil/projects/pmm/ (Monet et al. 1998). 

The principal goal of PMM has been to produce a digital sky survey for a
variety of scientific purposes and as a resource for the community.  The
first catalog of point sources measured from the original POSS~I, the
SRC-J and ESO-R survey (the latter two for the non-overlapping southern
sky) is called USNO-A.  At the time of this writing, all plates accepted
for POSS~II have also been scanned with the PMM.  Many second-generation
R-band plates taken by the UK Schmidt telescope have also been scanned.
The long term goal is to produce the USNO-B catalog, which will extend
the USNO-A catalog in key areas: providing star/galaxy differentiation
information and proper motions for point sources detected at both
Palomar epochs.

The first POSS~II plates were obtained in 1986, and they continue to be
taken up to the time of this writing.  Thus the time baseline for
measuring the proper motion of stars exceeds 30 years, and may extend to
45 years for some areas of the sky.  For comparison, the Luyten-Palomar
time baseline was 10-15 years.  Clearly, the search radius for detecting
the migrations of stars having large motion must be increased
correspondingly in matching POSS~II with POSS~I stellar positions.  This
makes the task of finding the correct match more difficult than that
faced by Luyten.  Thus, while it is easy to forecast that many more
proper motion stars will be catalogued in USNO-B than exist in current
catalogs for motions near or below the limits of those catalogs
(0.2$^{''}$~yr$^{-1}$ for Luyten), it may be difficult even with a new
generation of software and hardware to find the stars of very high
motion.

Given the high standards for acceptance of POSS II plate material, a
significant fraction of the 894 fields have been imaged more than once
in at least one of the B, R or I bands. A proportion of the latter
plates have been rejected for cosmetic reasons (aircraft trails,
background density variations, small-scale photographic defects) or
because the image quality (primarily due to mild astigmatism) fails to
meet survey criteria. Many of those plates remain useful
scientifically, particularly for analyses requiring positional (rather
than photometric) data for point sources, such as the astrometric study
described here.

The purpose of this paper is to assess the completeness of surveys for
stars of large proper motion (notably the LHS) using higher quality
plate material spanning a smaller time baseline than used in the
original analyses.  Although only a small percentage of the total sky is
covered, we also had the chance to find much more easily than Luyten any
stars in these fields with proper motions $\geq$ 2.5$^{''}$~yr$^{-1}$.
We adopt a minimum value of 0.4$^{''}$~yr$^{-1}$, placing us below the
lower bound of the LHS, but we are interested primarily in motions
appreciably larger than that.  We place an additional arbitrary
constraint on the survey F plate brightness, restricting ourselves to
objects fainter than F = 14.  Other proper motion surveys for brighter
stars (eg. from the Lowell Observatory) lend greater confidence in the
completeness level for the brighter stars.  The goal of our analysis was
therefore (1) to find new stars that Luyten missed with proper motions
larger than 0.4$^{''}$~yr$^{-1}$, and fainter than F=14, and (2) to test
the completeness of our own methods by finding the percentage of
catalogued proper motion stars that we recover.  This presumably also
tests our efficiency in finding proper motion stars missed in especially
the LHS survey, and in testing the completeness of the LHS.  Our data
set and procedures are described in Section~2, and the results presented
in Section~3.  The implications for the completeness of prior catalogs,
and the promise of the POSS~II data for the discovery of new proper
motion stars, are discussed in Section~4.

\section{PROCEDURE}

For this project we selected fields where rejected red (POSS~II IIIaF,
hereafter F$_{rej}$) plates of good quality were available, as well as
accepted plates for the blue and infrared bands.  The reasons for this
choice included the fact that stars of large proper motion are generally
cool, that more such pairs were available than for blue (POSS~II IIIaJ)
pairs, and that Luyten measured proper motions using the red (E) plates
of both his epochs.  A drawback of the sample is that the epoch
differences between the rejected and accepted F plates, and between these
and the accepted blue (J) and and infrared (IV-N or N) plates are random
and can range from essentially zero to as long as 10 years.

Rejected F plates from 140 fields have been scanned with the PMM at
Flagstaff.  In our initial analysis we concentrated on fields at high
Galactic latitude ($|b| >$ 20$^o$) with full four-plate sets (POSS~II
J, F, F$_{rej}$ and N). In addition, we required that the set of four
plates provide a suitable spread in epoch distribution: specifically,
no epoch difference between any two of the four plates could be less
than 1.5 years.  Any proper motion object found thus was required to
show a consistent pattern of motion (the same amount and direction)
over four different epochs. The reason for this restriction is the
prevalence of ``false'' detections of apparent point sources in the
halos of very bright stars, in their diffraction spikes, within
galaxies, and between double stars.  The vast majority of these are
filtered out by requiring the four-plate solution for motion to be
consistent.  It was feared that this stringent requirement might result
in the inability to find some ``real'' objects owing to positional
measurement errors.  It was hoped, however, that by maintaining the
four-plate requirement, the number of spurious detections would be
reduced to a manageable number, excludable by inspection of a digitized
POSS server through the Internet.  (Given the very large number of
fields so inspected, this procedure was more practical than downloading
our own pixel database stored on many tapes.)  The imposition of these
two requirements reduced the available fields from around 140 to 35.
This corresponds to roughly 5\% of the Luyten POSS~I fields.  The plate
centers of these 35 POSS~II fields are listed in Table~1.

To ascertain the degree to which we inhibit our ability to recover
real proper motion objects by requiring that there be consistent
measurements on all four plates spanning the blue to infrared bands,
we repeated the procedure using two combinations of just three plates.
We omitted the N plate in one set, to test how much that (generally)
lower quality plate affects our completeness.  In the other set, we
dropped the requirement of a J plate detection to determine the number
of possibly-cool objects measured only on the F and N plates.

A more detailed description of the procedure for matching point
sources on the several plates used for each field, and the measurement
of consistent proper motion candidates between pairs of plates is
given in Appendix~A.

\section{RESULTS}

\subsection{New proper motion stars} 

In the 35 fields for the four-plate solution, there were 50 objects
classified as stars by the software that met our motion and magnitude
constraints but with no corresponding entry in the Luyten or Giclas
catalogs.  Of these, 31 were objects misidentified as point sources by
the PMM but which visual inspection showed to be galaxies, close
doubles, or diffraction spikes associated with very bright stars usually
having substantial proper motion.  Nine objects showed no visible motion
between the digitized versions of POSS~I and POSS~II.  That is, there
could be no confirmation of a real proper motion star in the field down
to the POSS~I magnitude limit.  Three objects proved to be known,
catalogued stars, but the identification was missed by the software.
The remaining seven objects appear to be genuinely new proper motion
objects, one of which is actually a common proper motion pair.  These
eight new proper motion stars, their J2000 positions, proper motions and
position angles, and the four photographic magnitudes, are given in
Table~2.  Finding charts are not displayed here, since each star is
easily accessible via a POSS web server
(http://www.nofs.navy.mil/data/FchPix/cfra.html).  Proper motion may
also be confirmed by comparison with the POSS~I and POSS~II fields.
  
Nine new proper motion stars from the two three-plate solutions are
given in Table~3.  The first two objects were found from the solutions
lacking the J plates, and the last seven lacked the N plates.  Not
surprisingly, a much higher fraction of ``mistakes'' occurred than for
the four-plate solution (diffraction spikes, galaxies, doubles and
blended images, etc.) and again had to be eliminated by visual
inspection of the digitized POSS~I and POSS~II fields. 

Of the 17 new proper motion stars of Tables~2 and 3, 10 are below the
0.5$^{''}$~yr$^{-1}$ limit of the LHS.  While we used a generously low
cutoff proper motion of 0.4$^{''}$~yr$^{-1}$ in this exercise, the main
interest is in the completeness for finding those of higher motion, such
as above 0.6$^{''}$~yr$^{-1}$ used previously for defining samples of
white dwarfs (Dahn, Monet \& Harris 1989, Liebert et al. 1999) and halo
subdwarfs (Dahn et al. 1995).  Four such objects are listed in Table~2,
and only two in Table~3 had motions exceeding 0.6$^{''}$~yr$^{-1}$.  In
the remainder of this section, we discuss some of these stars
individually.  Table~4 presents some broadband optical photometry we
were able to obtain for the new proper motion stars in Table~2.

POSS~15:00:03.51~+36:00:30.5: Inspection of the POSS~II image shows a
clear point source at the J2000 position.  A similar inspection of the
POSS~I shows a marginally-visible candidate at the expected separation
and position angle. The photometry in Table~4 indicates the object is
fainter than Luyten's POSS~I and second epoch detection limit, and is
certainly below the LHS completeness limit. A spectrum with the Hale 5~m
double spectrograph (Oke \& Gunn 1982) was obtained on 1998 May, and the
blue and red spectra are shown in Fig.~1ab.  Atmospheric features in the
red part of the spectrum have not been removed.  No stellar features are
detected in the spectrum.  In particular the absence of an H$\alpha$
line indicates that the spectral type, at least in the red, is that of a
very cool DC white dwarf.  A parallax is being measured with the
$1.55\rm\,m$ Strand Telescope, and the preliminary result confirms the
spectroscopic suggestion.  After just over one year of observation, the
absolute parallax is $15.65\pm 1.0\rm\,mas$, and the absolute V magnitude
is $15.33\pm 0.14$.  This result is consistent with the M$_V$ estimatable
from the BVI colors (Bergeron, Wesemael \& Beauchamp 1995) and implies
that it is a cool white dwarf with a normal mass $\sim
0.60\rm\,M_{\odot}$.  Together with the proper motion, this indicates a
tangential velocity of $275\pm 18\rm\,km\,s^{-1},$ showing that the
object is a halo star.

POSS~15:30:55.62~+56:08:56.4/15:30:56.51~+56:08:52.0: This is a resolved
double whose components are of nearly equal brightness.  Our software
paired the components wrongly and produced a single proper motion star
with 1.438$^{''}$~yr$^{-1}$.  Inspection showed that both components
move together and the remeasured values were 0.738$^{''}$~yr$^{-1}$ with
position angle $56.5^{\circ}$.  Luyten (1981) wrote ``The
machine-processed plates lack virtually all double stars with nearly
equal magnitudes and separations less than 10 arc seconds'' because it
ignored ``nonstellar'' objects. Spectra were obtained of both stars on
1999 July 17 U.T. with the LRIS spectrograph on Keck~II (Oke et
al. 1995), used in the configuration described in Kirkpatrick et
al. (1999). These spectra are shown in Fig.~2, fluxed on an F$_\lambda$
scale.  Strong features due to CaH near 6900\AA\ demonstrate that each
object is an extreme subdwarf M, logical spectral counterparts for stars
of large proper motion.

POSS~21:01:04.18~+03:07:05.1: The POSS~I image appears blended with a field
star, though our POSS~II epochs have clean images.  This presumably
explains why this object was not in the Luyten catalog.  Since its
colors are quite red, it is likely to be a late M dwarf possibly 
within 20 pc.

POSS~00:01:25.72~+28:25:20.5 and 16:17:50.84~+19:05:43.2: These are the
only two new proper motion stars with $\mu\ >$ 0.5$^{''}$~yr$^{-1}$
found from three-plate solutions that appear clearly on both digitized
POSS epochs.  At magnitudes brighter than 19 in all bands, there is no
apparent reason they fail to appear in a Luyten catalog.  We have not
yet obtained photometric, spectroscopic or astrometric observations.

\subsection{Recovery efficiency of catalogued proper motion stars} 

To evaluate the completeness of our own survey, we analyzed the fraction
of previously catalogued objects that the PMM successfully identified.
There are a total of 230 fast ($\mu\geq 0.4^{''}\,\rm yr^{-1}$) and
faint (magnitude 13.0 in some band or fainter) catalogued proper motion
stars in these 35 fields.  Of these 230 stars, 216 were recovered, or
94\%.  Of the remaining 14 objects, six stars are sufficiently near the
edge ($\leq 1\rm\,cm$ or $\approx 10\rm\,arcmin$) of their respective
plates, that there may be either larger errors in determining the
astrometry or sufficient deviation among the four plate centers to put
the star off the field of at least one of the plates.  The remaining
eight stars -- less than 3.5\% of the catalogued stars -- appear to have
been missed by the software for various reasons, some explainable and
others not.  (One object was in the flare of a bright star, one was
nearby a bright star, two were perhaps too bright for the survey, two
were blended, and two had no obvious explanation.)  The effective survey
area for each plate is $34\times 34\rm\,cm^2$ or $6.35\times 6.35\rm\,
deg^2$ minus the occluded area of the density spots ($4.5\times
5.7\rm\,cm^2$ or $0.894\rm\,deg^2$), so that our 35 plate search covered
some $1409-31=1378\rm\,deg^2$ , or 3.3\% of the sky.

\section{DISCUSSION} 

Luyten processed by machine or eye some 804 POSS~I fields, or roughly
$28000\rm\,deg^2$ at significant Galactic latitudes.  Some 160 low
Galactic latitude fields were not analyzed, and he estimated (Luyten
1979) that about twice that area of sky was inaccessible to Palomar
observation.  The LHS Catalog lists 3587 stars with proper motions above
0.5$^{''}$~yr$^{-1}$.  In this experiment on some 1378 square degrees --
$\sim$3\% of the sky or $\sim$5\% of the fraction measured by Luyten --
we discovered six new stars of such large proper motion.  This
experiment also offered the possibility of finding very high proper
motion stars, such as nearby halo white dwarfs above Luyten's
2.5$^{''}$~yr$^{-1}$ search radius.  However, no such star was
found. Our results suggest that a similar POSS~II survey of all 804
Luyten fields might add a modest 6/0.05 or $\sim$120 new ``LHS'' stars
plus doubles, or $\sim$130 objects if our $>$90\% success rate in
recovering catalogued LHS stars in these fields is indicative of our
overall completeness.  The proper motions of the new stars appear
consistent with the expected $\mu^{-3}$ distribution.  If our yield of
new stars is extrapolated to 2.5$^{''}$~yr$^{-1}$, only about two new
stars of larger motion would be predicted for the entire sky.

We have also experienced firsthand a few circumstances that might have
caused Luyten to have missed a motion star -- superposition with another
star during one epoch, close common proper motion pairs, and of course
the modest limiting magnitudes of POSS~I.  Much ado has been made of the
fact that Luyten missed the large proper motion star WD0346+246, which
is claimed to be a very cool halo white dwarf (Hambly, Smartt and
Hodgkin 1997; Hodgkin et al 2000).  However, this object is very faint,
similar to the new, probable halo white dwarf
POSS~15:00:03.51~+36:00:30.5; the second epoch Luyten exposures may have
been too short to detect either star.

While this was a painstaking experiment, we do not claim an accurate
value of the completeness of the LHS Catalog.  However, if the LHS were
only $\sim$60\% complete to a red magnitude of R$_L$=18.5 (Hambly et
al. 1997; Flynn et al. 2000), over one hundred new proper motion stars
might have been found in this exercise.  Our work indicates instead that
the completeness fractions reported by Hanson (1979) and Dawson (1986)
(80-90\%) are not serious overestimates.  A similar conclusion is
implied by the analysis of the cumulative distribution of proper motions
for an enlarged group of LHS-selected white dwarfs (Liebert et al. 1999;
Harris et al. 2000).

The discrepancy between the completeness of Luyten's catalogs found in
this paper and that found by Flynn et al. (2000) appears significant.
What might cause such a discrepancy?  We have reviewed the clever
argument made by Flynn et al., and we have confirmed their results
(their Fig. A1).  One assumption made in their analysis that ``the
number density of stars does not change appreciably on scales equivalent
to a distance modulus of 0.5 mag'' is not strictly valid at high
Galactic latitudes because their space density drops away from the
Galactic disk.  Therefore, the cumulative effect of half-magnitude steps
from magnitude 13 to 18 in the Flynn et al. analysis might give a
spuriously-low completeness.  In fact, a repeat of the analysis on
subsamples of the NLTT is shown in Figure 3.  The greater
``incompletness'' found at high latitude could be caused by this drop in
stellar density away from the disk. 

A further exploration of the effect of density variations has been made
with a Monte-Carlo kinematic model of the disk.  Using realistic values
of the velocity ellipsoid, the luminosity function, and the space
density profile of red dwarfs, we find that the NLTT has 5-10\% fewer
stars at $|b| >$ 55$^o$ than would be the case with a uniform space
density. (The exact fraction depends on input parameters, particularly
on the assumed velocity dispersion.) Therefore, a portion of the
incompleteness found in the Flynn et al. analysis can be explained by
this effect.  Other effects, such as crowding and reddening at low
latitudes, may also be important factors. 
 
Our findings in this paper constrain hypotheses about the possible
contribution of white dwarfs to the dynamical or ``dark'' halo, such as
Alcock et al. (2000) and Ibata et al. (2000) among others have
suggested.  An optimistic assumption is that the peak in such a halo
white dwarf distribution occurs near M$_V$ = +17 to 18.  Flynn et
al. (2000) conclude that, if Ibata et al.''s (2000) claim to have
discovered several halo white dwarfs in the Hubble Deep Field were
correct, that several times that number of similar stars should have
been found in the greater search volume accessed by existing
ground-based catalogs, principally the LHS.  This might have been
possible if the LHS were substantially incomplete. With our finding of a
substantially greater completeness for the LHS, Flynn et al.'s
conclusion is strengthened.  Likewise Chabrier's (1999) models also
suggest that dozens of such stars might be detectable out to a distance
of 15 pc that even a modest V=18 magnitude limit might permit.  Of
course Hansen (1998), Chabrier (1999) and others have shown that --
depending on their mass distribution and outer layer compositions -- the
oldest halo white dwarfs may be much fainter than M$_V$ = +17-18 and not
necessarily detectable to LHS limiting magnitudes.

\acknowledgments { } JL and MDF acknowledge the hospitality of the U.S.
Naval Observatory during which most of this work was done.  We thank
Conard Dahn for useful discussions and comments on this manuscript.  We
also thank the referee, Bob Hanson, for several helpful comments and
criticisms of the paper.
\appendix

\section{Procedures Used for Identification and Measurement of Proper 
Motion Stars from Several Plates} 

The Palomar Observatory Sky Survey II (POSS-II) original plate material
has been scanned by the PMM.  Three emulsions (red, called ``SF''; blue,
called ``SJ''; and near infrared, called ``SN'') comprise POSS-II.  All
accepted POSS-II plates available have been scanned.  In addition, SF
and SJ plates that were rejected (owing to the presence of airplane
lights, for instance) have also been scanned.  Fields for which the
epoch difference was at least 1.5 years between any two plates (from
among the SF, SJ, SN, and either of the rejected plates -- ``RF'' or
``RJ'') were selected as eligible for finding high proper motion
stars, at Galactic latitude ($|b|\geq\ 20^{\circ}$).  The use of
POSS-II means that plate material with more than three epochs is
available, all of which have the same pointing center (to within a few
arcminutes).  The astrometric reduction is thus considerably simplified.

The procedure to identify objects of high proper motion begins with
removing duplicate detections from the raw object list.  The
peak-finding and blob-splitting software that is part of the real-time
PMM image processing sometimes finds (spurious) multiple peaks at the
position of a single star.  Only the brightest object within a
$1\rm\,arcsec$ radius is kept in the list for further processing.
Postprocessing software using oblique decision trees works with the
computed image parameters (the image moments, shape parameters, etc.)
to deduce a classification for the object (star or galaxy) on a sliding
scale from 0--11 (zero meaning almost certainly ``galaxy'' and 11
meaning almost certainly ``star'', with the transition of maximal
uncertainty being between five and six).

The list of objects surviving the removal of duplicate detections (for
each plate) is reduced with respect to the ACT (Urban, Corbin, \& Wycoff
1997).  The object positions are modified according to the SLALIB
pincushion distortion for the 48-inch Oschin Schmidt telescope, plus the
so-called ``taffogram'' (the systematic astrometric residual as a
function of plate position after removal of the SLALIB pincushion).  Ten 
coefficients are used in the astrometric solution to derive $(\xi,
\eta)$ (the tangent plane coordinates) according to the ACT.  No
magnitude term was used in the astrometric solution.  There are now
four lists of $(\xi, \eta)$ coordinates, one for each plate.

Using the four lists of $(\xi, \eta)$, pairings of matched objects are
made between the SF coordinate list and the other three coordinate
lists.  An object in the SJ, SN, or RF list that is found within a
radius $r_{\rm del} = \mu_{\rm min}\Delta t_{\rm SF-X}$ of an SF object
is marked for deletion in both lists.  Here, $\mu_{\rm min} =
0.2\rm\,arcsec\,yr^{-1}$ is the lower bound on proper motion to which
our program is sensitive.  This bound is based on the mean error of our
positional measurements for a single object, which is about
$0.3\rm\,arcsec$, and the minimum acceptable epoch difference between
plates, which is $1.5\rm\,yr$.  The time interval $\Delta t_{\rm SF-X}$
is the epoch difference between the SF plate and paired plate (``X'')
being examined.  Following the identification of all pairs of matching
objects between the SF plate and the other three plates, those objects
are all removed from the object lists.  Only the surviving objects are
subjected to further processing.

The remaining objects are examined for consistent linear proper motion
in the next phase of processing.  For each object on the surviving SF
list, the RF list is examined and all objects within a radius of
$r_{\rm pm} = \mu_{\rm max}\Delta t_{\rm SF-RF}$ are identified.
Here, $\mu_{\rm max} = 6\rm\,arcsec\,yr^{-1}$ is the maximum proper
motion we choose to investigate.  The number of proper motion
candidates increases as the square of this radius, so limiting it to a
reasonable number results in saving computing time and limiting the
false detection rate.

The lists of all SJ objects and all SN objects that are within $r_{\rm
pm}$ of each SF-RF pair are similarly compiled.  If there is no SJ or
SN object within $r_{\rm pm}$, that pair of SF/RF objects is removed
from further consideration.

Following construction of the lists of SF-RF-SJ-SN candidate lists,
all possible combinations of objects (choosing one object at a time
from the each of the RF, SJ, and SN lists associated with an SF
object) are examined for consistency with a linear proper motion.  
A proper motion is predicted from the SF and RF positions:  
\begin{equation}
\mu_x = (x_{\rm RF} - x_{\rm SF})/(t_{\rm RF} - t_{\rm SF}) 
\end{equation}
is the predicted $x$-component of the proper motion, for instance.
Then the position for, say, the SJ object is predicted using the
interpolation formula
\begin{equation}
x_{\rm SJ} = x_{\rm SF} + (t_{\rm SJ} - t_{\rm SF})\mu_x.
\end{equation}
There is a tolerance $\sigma_{\rm SJ}$ around the predicted position
$x_{\rm SJ}$ whose size is
\begin{equation}
\sigma_{\rm SJ} = \sigma_{\rm SF}\left[ 1 + 2\frac{t_{\rm SJ} - t_{\rm
SF}}{t_{\rm RF} - t_{\rm SF}}\right]^{1/2}.
\end{equation}
We take $\sigma_{\rm SF}\equiv 1\rm\,arcsec$.  An object in the SJ list
that is within a radius $\sigma_{SJ}$ of the predicted position
$(x_{\rm SJ}, y_{\rm SJ})$ remains a viable proper motion candidate.  A
similar procedure is then used for the SN surviving object list.  The
final list of proper motion candidates is constructed from objects such
that there is a detection on all four plates satisfying the linear
proper motion consistency criterion described here.

\vfill
\eject

\clearpage

\begin{figure}
\begin{center}
\figurenum{1a}
\epsfxsize=8cm
\epsffile{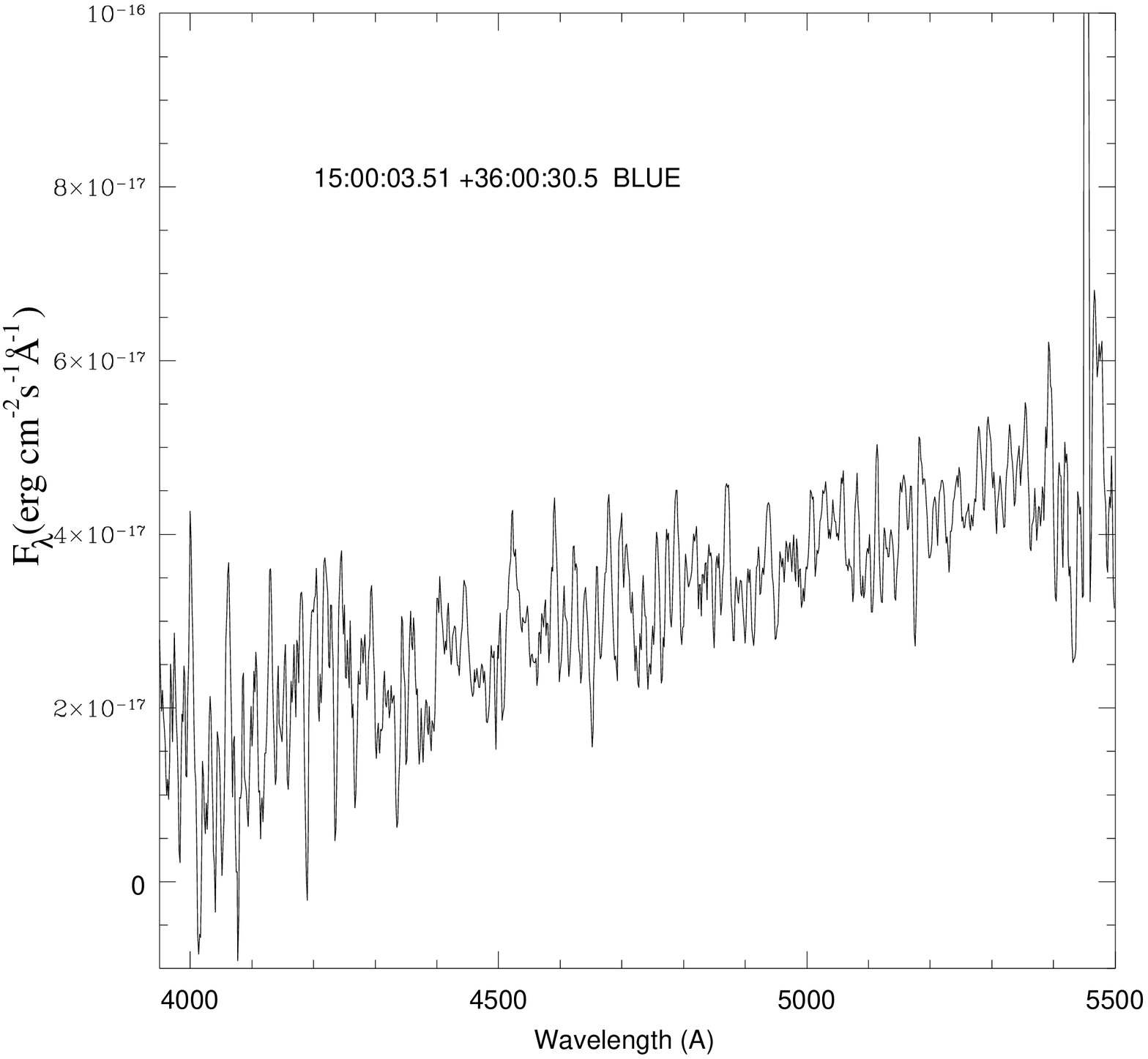}
\caption{Blue spectrophotometry of the new halo white dwarf with the Hale
(Palomar) 5~m and double spectrograph.  F$_{\lambda}$ in units of
ergs~cm$^{-2}$~s$^{-1}$~\AA$^{-1}$ is plotted against the wavelength in
\AA. }
\end{center}
\end{figure}

\begin{figure}
\begin{center}
\figurenum{1b}
\epsfxsize=8cm
\epsffile{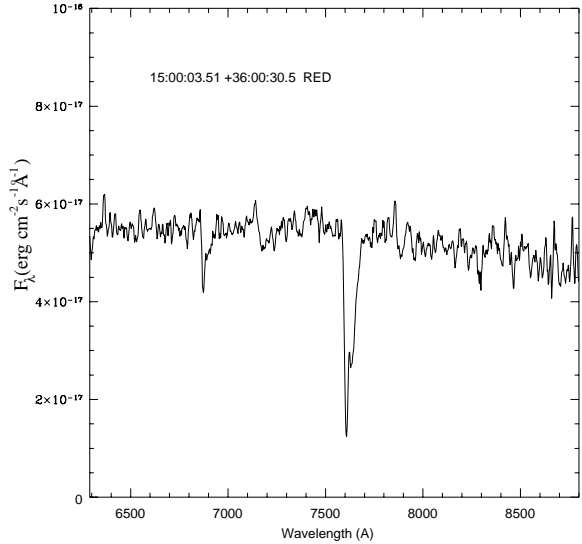}
\caption{The red half of the spectrum of the halo white dwarf, in the 
same units. Terrestrial atmospheric features in the red have not been 
removed. The spectrum is classified DC.}
\end{center}
\end{figure}
  
\begin{figure}
\figurenum{2}
\plotone{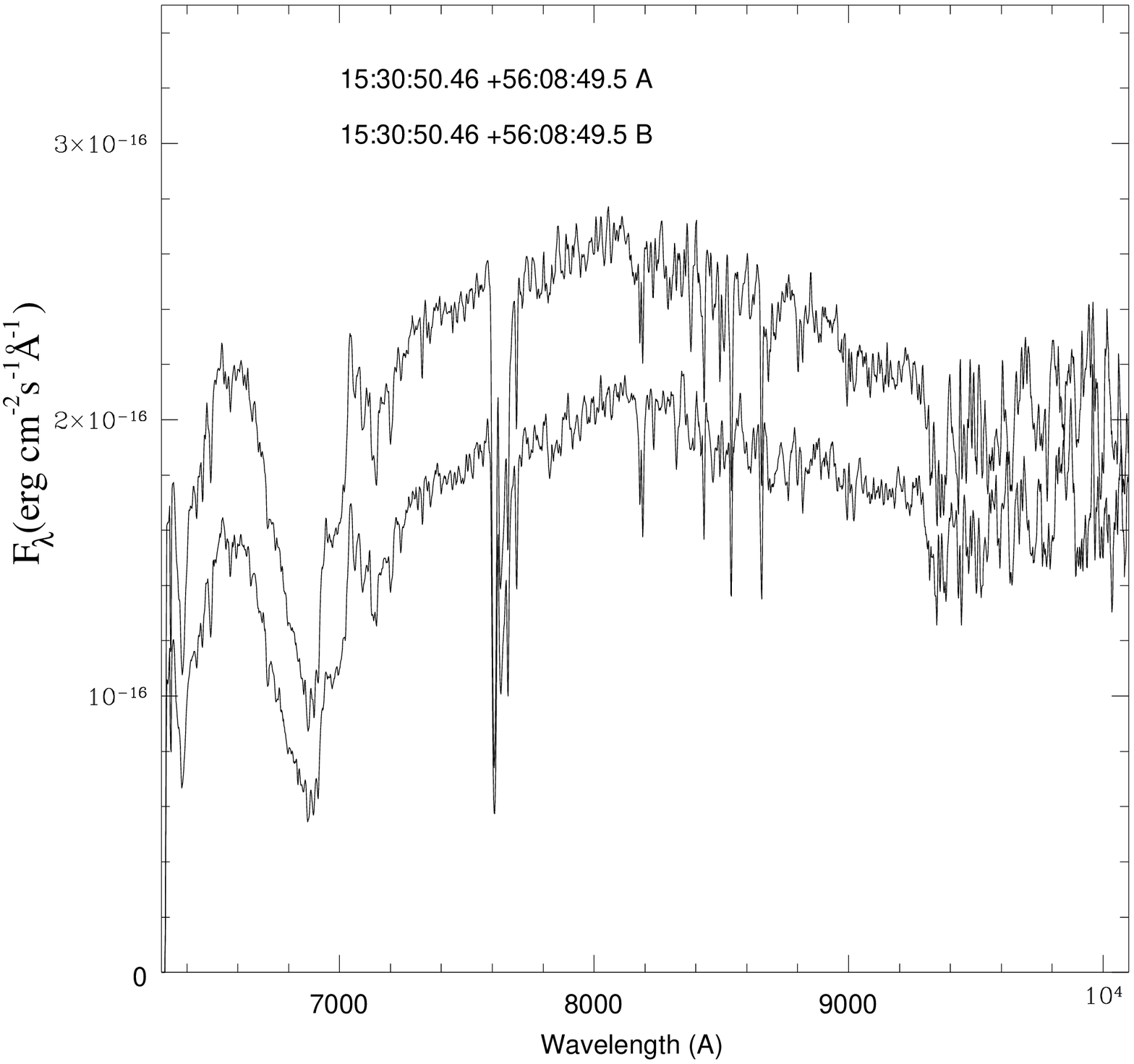}
\caption{Spectra of both extreme subdwarf M of the common proper motion
binary, taken with Keck~II and the LRIS spectrograph, with the same
units as Fig.~1.}
\end{figure}

\begin{figure}
\figurenum{3}
\plotone{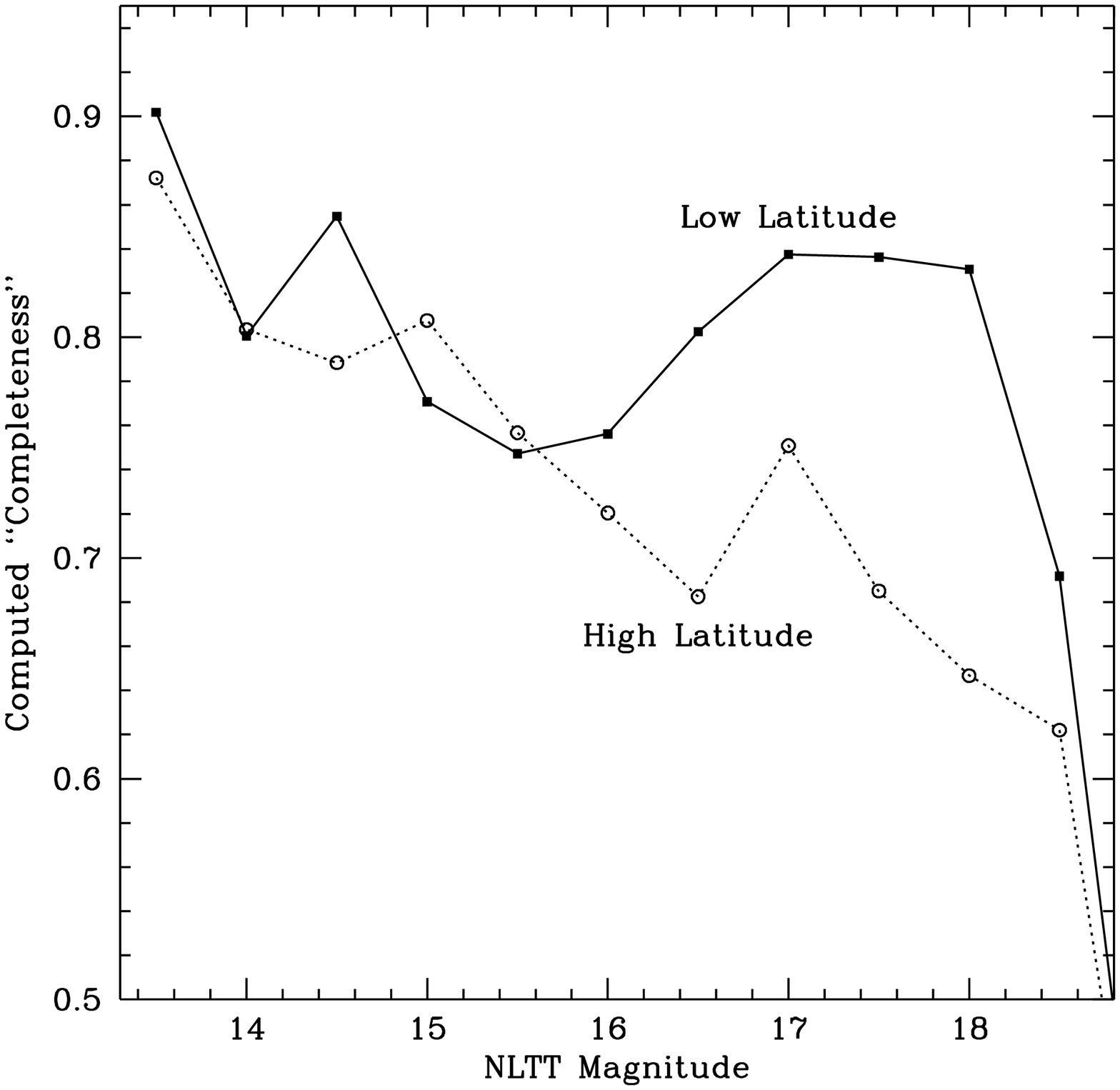}
\caption{Completeness of NLTT Catalog relative to R$_L$ = 13 in
Completed Palomar Region of the sky as used by Flynn et al. (2000), but
divided into high Galactic latitude
(55$^{\circ}\leq|b|\leq$90$^{\circ}$, open circles, dotted lines) and
low (15$^{\circ}\leq|b|\leq$35$^{\circ}$, filled squares, solid lines).}

\end{figure}

\clearpage 

\begin{deluxetable}{ccccc}
\tablenum{1} 
\tablewidth{17cm} 

\tablecaption{Plate Centers for the Thirty-five Fields}

\tablehead{
\colhead{Field} &\colhead{RA(J)} &\colhead{Dec(F)} &\colhead{RA(F)} 
&\colhead{Dec(F)}}

\startdata
140 & 18:22:38 & 60:01:32 & 18:22:38 & 60:01:35 \nl 
163 & 07:45:58 & 54:52:54 & 07:45:58 & 54:52:41 \nl 
177 & 15:25:19 & 54:49:29 & 15:25:25 & 54:49:40 \nl 
218 & 12:32:22 & 49:43:27 & 12:32:22 & 49:43:32 \nl 
223 & 15:01:37 & 49:48:12 & 15:01:43 & 49:48:24 \nl 
333 & 17:21:38 & 39:57:04 & 17:21:38 & 39:57:05 \nl 
336 & 18:39:38 & 40:02:46 & 18:39:39 & 40:02:47 \nl 
387 & 15:14:00 & 34:48:48 & 15:13:58 & 34:48:54 \nl 
408 & 23:38:28 & 35:16:36 & 23:38:28 & 35:16:33 \nl 
410 & 00:25:36 & 30:16:28 & 00:25:38 & 30:16:38 \nl 
411 & 00:48:42 & 30:16:17 & 00:48:42 & 30:16:20 \nl 
471 & 23:48:31 & 30:16:37 & 23:48:31 & 30:16:40 \nl 
477 & 01:52:47 & 25:14:38 & 01:52:48 & 25:14:45 \nl 
482 & 03:42:59 & 25:09:28 & 03:43:00 & 25:09:28 \nl 
509 & 13:36:26 & 24:44:55 & 13:36:20 & 24:44:45 \nl 
537 & 23:52:32 & 25:16:38 & 23:52:32 & 25:16:37 \nl 
539 & 00:23:36 & 20:16:36 & 00:23:36 & 20:16:31 \nl 
545 & 02:29:48 & 20:13:15 & 02:29:48 & 20:13:11 \nl 
584 & 16:08:11 & 19:52:08 & 16:08:11 & 19:52:05 \nl 
600 & 21:44:20 & 20:13:46 & 21:44:18 & 20:13:36 \nl 
601 & 22:05:21 & 20:14:32 & 22:05:21 & 20:14:33 \nl 
618 & 03:42:48 & 15:09:26 & 03:42:49 & 15:09:30 \nl 
650 & 14:22:24 & 14:46:11 & 14:22:22 & 14:46:27 \nl 
673 & 22:02:23 & 15:14:21 & 22:02:24 & 15:14:28 \nl 
711 & 10:42:42 & 09:44:26 & 10:42:38 & 09:44:06 \nl 
760 & 03:02:37 & 05:11:42 & 03:02:38 & 05:11:41 \nl 
792 & 13:42:36 & 04:45:06 & 13:42:32 & 04:44:49 \nl 
800 & 16:22:29 & 04:52:55 & 16:22:28 & 04:53:00 \nl 
814 & 21:02:29 & 05:11:48 & 21:02:28 & 05:11:48 \nl 
819 & 22:42:31 & 05:15:43 & 22:42:32 & 05:15:40 \nl 
827 & 01:22:33 & 00:15:37 & 01:22:33 & 00:15:35 \nl 
835 & 04:02:33 & 00:08:16 & 04:02:32 & 00:08:11 \nl 
885 & 20:42:34 & 00:10:49 & 20:42:31 & 00:10:41 \nl 
886 & 21:02:33 & 00:11:51 & 21:02:33 & 00:11:48 \nl 
890 & 22:22:33 & 00:15:07 & 22:22:33 & 00:15:05 \nl 

\enddata
\end{deluxetable}

\begin{deluxetable}{ccccccccc}
\tablenum{2} 
\large 
\tablewidth{17cm}

\tablecaption{New Motion Stars from Four Plates}

\tablehead{
\colhead{J2000} &\colhead{$\mu$} &\colhead{$\sigma$} &\colhead{PA} &
\colhead{$\sigma$} &\colhead{F} &\colhead{F$_{rej}$} &\colhead{J} &\colhead{N}}

\startdata
02:35:36.80~+22:33:02.8 & ~0.40 & ~0.05 & ~117 & ~07 & ~16.79 & ~16.58 & ~18.71 & ~15.17 \nl 
15:00:03.51~+36:00:30.5 & ~0.97 & ~0.05 & ~231 & ~03 & ~18.90 & ~18.07 & ~17.79 & ~20.24 \nl 
15:30:55.62~+56:08:56.4 & ~0.74 & ~0.07 & ~057 & ~05 & ~17.17 & ~17.14 & ~17.92 & ~16.51 \nl 
15:30:56.51~+56:08:52.0 & ~0.74 & ~0.07 & ~057 & ~05 & ~17.17 & ~17.14 & ~17.92 & ~16.51 \nl 
18:43:55.83~+41:55:43.3 & ~0.51 & ~0.05 & ~203 & ~06 & ~14.74 & ~14.68 & ~15.10 & ~15.61 \nl  
21:01:04.18~+03:07:05.1 & ~1.04 & ~0.05 & ~092 & ~03 & ~17.16 & ~16.96 & ~18.82 & ~15.35 \nl
22:04:04.48~+13:45:18.5 & ~0.49 & ~0.06 & ~061 & ~07 & ~16.27 & ~15.25 & ~17.25 & ~15.49 \nl 
23:47:56.35~+29:42:23.6 & ~0.44 & ~0.07 & ~114 & ~09 & ~17.28 & ~17.25 & ~18.57 & ~16.04 \nl 

\enddata

\end{deluxetable}

\begin{deluxetable}{cccccccc}
\tablenum{3} 
\large 
\tablewidth{17cm}

\tablecaption{New Motion Stars from Three Plates}

\tablehead{
\colhead{J2000} &\colhead{$\mu$} &\colhead{$\sigma$} &\colhead{PA}
&\colhead{$\sigma$} &\colhead{F} &\colhead{F$_{rej}$} &\colhead{J~or~N}}

\startdata
00:01:25.72~+28:25:20.5 & ~0.71 & ~0.08 & ~108 & ~07 & ~17.83 & ~18.47 & ~18.68 \nl
21:52:38.78~+22:07:28.7 & ~0.42 & ~0.11 & ~094 & ~14 & ~16.42 & ~16.24 & ~18.37 \nl
13:51:52.92~+04:41:46.4 & ~0.49 & ~0.07 & ~192 & ~08 & ~15.80 & ~15.78 & ~15.02 \nl 
13:40:56.68~+01:47:58.4 & ~0.42 & ~0.07 & ~160 & ~08 & ~14.12 & ~14.00 & ~14.29 \nl
16:19:10.95~+19:57:22.9 & ~0.48 & ~0.05 & ~300 & ~06 & ~14.05 & ~14.75 & ~14.25 \nl
16:17:50.84~+19:05:43.2 & ~0.65 & ~0.05 & ~224 & ~06 & ~17.22 & ~17.23 & ~18.21 \nl
22:13:17.42~+18:33:34.5 & ~0.47 & ~0.10 & ~220 & ~13 & ~16.08 & ~16.06 & ~15.72 \nl
22:05:33.10~+19:51:27.6 & ~0.40 & ~0.10 & ~186 & ~13 & ~15.61 & ~15.68 & ~15.51 \nl
23:58:21.88~+22:22:42.2 & ~0.43 & ~0.07 & ~170 & ~09 & ~16.65 & ~16.57 & ~16.00 \nl

\enddata

\end{deluxetable}

\begin{deluxetable}{ccccc}
\tablenum{4} 
\large 
\tablewidth{17cm}

\tablecaption{Photometry for New Motion Stars}

\tablehead{
\colhead{J2000} &\colhead{V} &\colhead{V-I} &\colhead{B-V} &\colhead{Nobs}}

\startdata
02:35:36.80 +22:33:02.8 & ~18.238 & ~3.665 &   ~-   & ~1 \nl 
15:00:03.51 +36:00:30.5 & ~19.354 & ~1.011 & ~1.063 & ~4 \nl 
15:30:55.62 +56:08:56.4 & ~18.500 & ~2.186 &   ~-   & ~2 \nl 
15:30:56.51 +56:08:52.0 & ~19.056 & ~2.306 &   ~-   & ~2 \nl 
18:43:55.83 +41:55:43.3 & ~16.574 & ~2.001 & ~1.452 & ~2 \nl  
21:01:04.18 +03:07:05.1 & ~18.734 & ~4.395 & ~2.02  & ~2 \nl 
22:04:04.48 +13:45:18.5 & ~17.167 & ~3.176 & ~1.766 & ~2 \nl 
23:47:56.35 +29:42:23.6 & ~18.591 & ~3.339 &  ~ -   & ~1 \nl 

\enddata
\end{deluxetable}

\end{document}